\documentclass{aa}  

\usepackage{graphicx}
\usepackage{txfonts}
\usepackage{lipsum}
\usepackage{subcaption}         
\usepackage{lscape}             
\usepackage{placeins}           
                                
\usepackage[colorlinks=true, linkcolor=blue, citecolor=blue, urlcolor=blue]{hyperref}
\usepackage{xcolor}
\usepackage{booktabs}
\usepackage{tabularx}
\usepackage{orcidlink}
\graphicspath{ {./plots} }

\begin{document}

   \title{The Birth of Minor-Body Populations in the $\beta$ Pictoris System I: The First 25 Myr.}

%
%
%
 
   \author{S. Torres\inst{1,2}\corrauth{santiago.torres@iac.es} \orcidlink{0000-0002-3150-8988}      
        \and 
        J.L. Gragera-M\'as \inst{3,4}\email{jlgragera@cab.inta-csic.es}\orcidlink{0009-0003-8081-2163}
        \and 
        A.J. Mustill\inst{5}\email{alexander.mustill@fysik.lu.se}\orcidlink{0000-0002-2086-3642}
        \and 
        E. Villaver\inst{1,2}\email{eva.villaver@iac.es}\orcidlink{0000-0003-4936-9418}}
         
   \institute{Instituto de Astrof\'sica de Canarias, 38200, La Laguna, Tenerife, Spain
   \and
   Universidad de La Laguna (ULL), Astrophysics Department, 38206, La Laguna, Tenerife, Spain
   \and
   Centro de Astrobiolog\'ia (CAB), CSIC-INTA, Camino Bajo del Castillo s/n, 28692, Villanueva de la Ca\~nada, Madrid, Spain \and Departamento de F\'{i}sica de la Tierra y Astrof\'{i}sica, Facultad de Ciencias F\'{i}sicas, Universidad Complutense de Madrid, E-28040, Madrid, Spain
   \and 
   Lund Observatory, Division of Astrophysics, Department of Physics, Lund University, Box 118, SE-221 00 Lund, Sweden}


\date{ }

 
  \abstract
  %
   {The discovery of the third known giant planet, $\beta$~Pic d, offers a new opportunity to investigate how planet--disk interactions build the minor-body architecture of the young Beta Pictoris system. We study the formation and redistribution of scattered-disk objects, falling evaporating bodies, a proto-Oort-cloud population, transitional interstellar objects, and interstellar objects. Using $N$-body simulations, we evolve a primordial planetesimal disk extending from 0.5 to 1000~au for 25~Myr under the perturbations of $\beta$~Pic b, c, and d, together with reconstructed stellar encounters from \textit{Gaia} DR3. The resulting evolution is strongly radius-dependent. The planets efficiently restructure the inner disk, while the outermost primordial population remains largely unchanged beyond $\sim325$~au. $\beta$~Pic c remains the main driver of the star-grazing population, showing that the falling-evaporating-body pathway persists after the inclusion of $\beta$~Pic d. At larger radii, $\beta$~Pic b and d dominate planetary scattering and produce a surviving disk edge near 50~au, consistent with the observed debris-disk truncation. Over 25~Myr, $8.74\%$ of the initial disk enters the scattered-disk population, $2.04\%$ reaches the proto-Oort-cloud region, and $6.44\%$ becomes unbound. Planetary stirring also generates a transient one-armed spiral among particles originating at $\sim60$--150~au, overlapping part of the observed ``Cat's Tail'' and the radial location of the prominent CO clump near 85~au. The reconstructed stellar encounters included here produce only weak perturbations, although the encounter catalog is complete only over the first ~2.5 Myr of the present epoch. The early evolution of the Beta Pictoris minor-body system is therefore dominated by its planetary architecture, offering a view of distinct minor-body populations while they are still being formed and redistributed. }


   \maketitle
   \nolinenumbers

\section{Introduction}
\label{intro}

Beta Pictoris (hereafter BP) provides an exceptional laboratory for investigating the dynamical interplay between planets, debris disks, and minor bodies. With an age of only $\sim20-25$ Myr \citep{Mamajek2014,Binks2014,Shkolnik2017,Miret-Roig2020,Esposito2020}, the system hosts an A6V-type star \citep{Gray2006}, a multiplanet architecture, and a massive debris disk with a complex and highly structured morphology. Until recently, BP was known to contain two giant planets: $\beta$ Pic c, a $\sim9\,M_{\rm Jup}$ planet orbiting at $\sim2.7$ au \citep{Lagrange2019,Nowak2020}, and $\beta$ Pic b, a $\sim10\,M_{\rm Jup}$ planet located at $\sim10$ au \citep{Lagrange2009,Lagrange2010}. These planets have long been regarded as the principal dynamical agents responsible for sculpting the observed disk and for driving the infall of star-grazing exocomets responsible for the well-known falling evaporating body (FEB) phenomenon in BP. More recently, independent evidence for a third giant planet, $\beta$ Pic d, has been reported by \citet{Sutlieff2026} and \citet{Gibbs2026}. Current estimates place $\beta$ Pic d at a semimajor axis of approximately $24$--$26$ au, with a mass of $\sim2.4\,M_{\rm Jup}$, although its orbital properties remain subject to further refinement. The presence of this additional outer planet substantially modifies the known architecture of the system and may play an important role in the early redistribution of planetesimals toward distant orbits.

This three-planet architecture is embedded within an extended and highly structured debris disk ranging from approximately $50$ au to at least $1000$ au \citep{Smith1984, Augereau2001, Apai2015, Ballering2016, Han2026}. In addition to this primary disk, the system also contains a secondary inner disk component inclined by $\sim5^{\circ}$ with respect to the main disk \citep{Golimowski2006,Apai2015} and extending out to  $\sim150$ au \citep[e.g.,][]{Milli2014,Millar-Blanchaer2015,Janson2021}. More recently, a prominent dust feature known as the ``Cat's Tail'' has been identified, extending in projection to approximately 230~au and likely associated with the secondary disk \citep[e.g.,][]{Rebollido2024}. The disk also exhibits a strongly asymmetric CO distribution, including a prominent clump near $\sim85$~au that has been interpreted as evidence for enhanced collisional activity \citep[e.g.,][]{Dent2014}. While the planetary orbits are broadly coplanar with the main disk, they remain misaligned with this secondary component.
     
The close dynamical coupling between the planets and the surrounding planetesimal population is also reflected in one of the most distinctive observational signatures of BP: its abundant exocomet activity. Exocomets were first detected through transient absorption features in high-resolution spectra \citep{Ferlet1987,Beust1990,Kiefer2014,Hoeijmakers2025}, interpreted as gas released by planetesimals on highly eccentric, star-grazing orbits as they transit the stellar disk, and also identified through broadband photometric variability \citep{Zieba2019}. These bodies, commonly referred to as falling evaporating bodies (FEBs), undergo rapid sublimation during close pericenter passages and produce the observed time-variable absorption features. High-resolution spectroscopy further indicates that the FEB population is dynamically diverse. \citet{Kiefer2014} identified two families of exocomets with characteristic pericentre distances of $\sim10\pm3\,R_\star$ and $\sim19\pm4\,R_\star$, while \citet{Heller2024} showed that the observed signals are consistent with highly eccentric bodies spanning a broad range of semimajor axes.

Several studies have investigated how planetary perturbations can drive FEB activity, showing that planetesimals may be delivered onto star-grazing orbits through mean-motion resonances (MMRs) \citep{Thebault2001,Beust2024,Rodet2024,Jaworska2026}. When only $\beta$ Pic b was known, models attributed FEB production primarily to low-order inner MMRs with this planet. The discovery and subsequent inclusion of $\beta$ Pic c significantly altered this picture by modifying the dynamical structure of the inner system. Resonant regions associated with $\beta$ Pic b become strongly destabilized, while higher-order MMRs with $\beta$ Pic c, located between approximately $0.6$ and $1.5$ au, emerge as more plausible pathways for generating FEBs \citep{Beust2024,Jaworska2026}. At larger distances, previous studies explored the possible role of additional planets in shaping the radial structure of the debris disk \citep{Lacquement2025}. The recent discovery of $\beta$ Pic d allows us to reassess this problem using the full three-planet architecture.

In this work, we investigate the early dynamical evolution of the BP planetesimal disk and assess how the newly established three-planet architecture shapes the disk during the first $25$ Myr. We examine the production of star-grazing bodies, the restructuring and truncation of the disk, and the scattering of planetesimals onto distant bound and unbound trajectories, thereby tracing the early emergence of the BP Oort cloud and a population of interstellar objects. Furthermore, our analysis builds upon the stellar-encounter reconstruction of \citet{JL_Torres2025}, who identified the past and future stellar flybys of BP using \textit{Gaia} DR3 astrometry \citep{Gaiacollaboration2018,GaiaEDR3Astrometry,GaiaDR3}. By combining the internal perturbations generated by the three planets with the reconstructed stellar environment, we determine which processes dominate the formation and early evolution of these dynamical populations. 

In Sect.~\ref{sec2}, we describe the numerical setup and initial conditions. In Sect.~\ref{sec3}, we examine the dynamical evolution of the planetary system and the resulting redistribution of the primordial planetesimal disk over the first $25$ Myr. Finally, we summarize our conclusions in Sect.~\ref{conclusion}.

\section{Numerical setup and data}
\label{sec2}

To investigate the dynamical evolution of the BP system and the formation of its minor-body populations, we used the N-body package \texttt{REBOUND} \citep{Rein2012,Rein2014,Tamayo2020}, which allows us to simultaneously account for planetary perturbations and stellar flybys. Our simulation includes BP, its three known planets, a disk of test particles, and a population of flyby stars; Galactic tides are not included in this work as they act on timescales longer than the current age of the system.

\subsection{Initial conditions}
\label{sec2.1}
We constructed the initial conditions using the three giant planets currently known in the BP system. The orbital parameters of $\beta$ Pic b and $\beta$ Pic c were adopted from the recent orbital solution of \citet{Macias2026}, consistent with previous determinations by \citet{Lacour2021}. For the recently discovered planet $\beta$ Pic d, whose orbit remains subject to refinement, we adopted a representative orbital solution reported by \citet{Sutlieff2026} and \citet{Gibbs2026}. The adopted planetary parameters are summarized in Table~\ref{table_planets}.

\begin{table}
\centering
\caption{Adopted planetary parameters for the BP system. The solutions for $\beta$ Pic b and $\beta$ Pic c were taken from \citet{Macias2026}, while $\beta$ Pic d is from \citet{Sutlieff2026} and \citet{Gibbs2026}. The listed orbital elements are the semimajor axis $a$, eccentricity $e$, inclination $i$, longitude of ascending node $\Omega$, argument of periastron $\omega$, and the epoch of periastron $\tau$, expressed as a fraction of the orbital period relative to the reference epoch.}
\label{table_planets}
\begin{tabular}{lccc}
\hline
Parameter & $\beta$ Pic c & $\beta$ Pic b & $\beta$ Pic d \\
\hline
Mass [$M_{\rm Jup}$]      & 9.03  & 10.0  & 2.4  \\
Radius [$R_{\rm Jup}$]    & 1.20  & 1.65  & 1.26 \\
$a$ [AU]                  & 2.688 & 10.03 & 26.0 \\
$e$                       & 0.235 & 0.109 & 0.19 \\
$i$ [$^\circ$]            & 1.06  & 1.01  & 1.00 \\
$\Omega$ [$^\circ$]       & 31.05 & 31.81 & 210.8 \\
$\omega$ [$^\circ$]       & 61.02 & 212.5 & 191.0 \\
$\tau$                    & 0.825 & 0.774 & 0.50 \\
\hline
\end{tabular}
\end{table}

We initialized a massless planetesimal disk composed of $20,000$ test particles with semimajor axes uniformly distributed between $0.5$ and $1000$ au. The inner boundary follows the initial conditions adopted by \citet{Beust2024} and \citet{Jaworska2026}, whose simulations identified dynamically active populations near $0.6$ and $1.5$ au that may contribute to the production of FEBs in the BP system. The outer boundary encompasses the observed extended disk, which reaches at least $\sim1000$ au, while also allowing us to follow the transport of planetesimals toward distant bound orbits.

The eccentricities were drawn from a uniform distribution over $0 \leq e \leq 0.05$, while the inclinations were randomly uniformly selected between $0^\circ$ and $2^\circ$. The argument of pericentre ($\omega$), longitude of the ascending node ($\Omega$), and mean anomaly ($M$) were uniformly distributed between $0$ and $2\pi$. These initial conditions are not intended to reproduce a particular physical surface-density profile. Instead, they define a simple reference ensemble for comparing dynamical outcomes across the full radial extent of the disk. The population fractions reported below therefore represent formation efficiencies relative to this adopted sampling and should not be interpreted as predictions for a specific primordial disk profile. In particular, because semimajor axes are sampled uniformly, the expected number of particles within a given radial interval is proportional to its width.

Particles were removed only following a physical collision with either the host star or one of the planets. No additional removal criteria based on distance from the host star, semimajor axis, or Hill-sphere crossings were imposed. Consequently, particles scattered onto highly eccentric or hyperbolic trajectories remained in the integrations, allowing us to follow their dynamical evolution and transitions between the populations.

The simulations were integrated for $25$ Myr, approximately corresponding to the present age of the BP system, and included the gravitational influence of the host star, the three known giant planets, and the stellar encounters identified by Gragera-Más et al. (in prep.). The initial conditions and main numerical parameters are summarized in Table~\ref{ic_sim}.

All simulations were performed using the \texttt{TRACE} integrator \citep{Lu2024,Pham2024}, a hybrid quasi-time-reversible scheme based on the framework of \citet{Hernandez_Dehnen2023}, with a timestep of $0.02$ yr and an output cadence of $5000$ yr. \texttt{TRACE} combines the efficiency of a symplectic treatment (\texttt{WHFast} \citep{Wisdom_Holman1991,Rein2015}) during regular orbital evolution with a more accurate treatment of close encounters (\texttt{IAS15} \citep{Rein2014}). This is particularly well suited to the BP system, where planetesimals may undergo repeated planetary encounters, be scattered onto highly eccentric trajectories, or evolve onto star-grazing orbits. The hybrid approach therefore allows us to resolve these dynamically active phases while retaining the computational efficiency required to follow thousands of particles over Myr timescales.

\begin{table}
\centering
\caption{Summary of the disk initial conditions and numerical setup adopted in this work.}
\label{ic_sim}
\begin{tabular}{ll}
\hline
Parameter & Value \\
\hline
Stellar mass & $1.75 M_\odot$ \\
Stellar radius & $1.52 R_\odot$ \\
Semi-major axis  & $0.5 \le a \le 1000$ au \\
Eccentricity & $0 \le e \le 0.05$ \\
Inclination & $0^\circ \le i \le 2^\circ$ \\
$\omega,\Omega,M$ & Uniform in $[0,2\pi]$ \\
Number of particles & 20,000 \\
Integration Time  & 25 Myr \\
Integrator & TRACE \\
Code & REBOUND \\
\hline
\end{tabular}
\end{table}

As discussed by \citet{Beust2024} and \citet{Jaworska2026}, the BP system presents a challenging configuration for direct N-body integrations. We therefore adopted $20,000$ massless test particles as a compromise between statistical sampling and computational cost. Our primary goal is to identify the dominant dynamical pathways leading to the different minor-body populations rather than to reproduce the detailed mass distribution of the primordial disk. The adopted model provides sufficient sampling to characterize the main dynamical mechanisms while remaining computationally tractable and reproducible. Increasing the number of simulated particles would mainly reduce sampling noise and make structures in the resulting populations better defined. It would not change the underlying dynamical result.

A further benefit of this moderate-resolution approach is the reduction in computational resources and energy consumption. Following the Green Algorithms framework\footnote{\url{https://www.green-algorithms.org/}} \citep{Lannelongue2021}, we estimate a carbon footprint of approximately $11~{\rm kg\,CO_2e}$ for the simulations presented in this work. Environmental considerations did not determine the scientific design of the calculations, but provided an additional motivation to avoid increasing the computational scale once the principal dynamical features were robustly identified.

\subsection{Close stellar encounters with Beta Pictoris}
\label{sec2.2} 

Stellar flybys can modify the orbital architecture of planetesimal populations by perturbing their eccentricities, inclinations, and perihelion distances over short timescales \citep[e.g.,][]{Heisler1986a,Pfalzner2018,Torres2023}. To account for the external dynamical environment of BP, we adopt the updated catalog of close stellar encounters compiled by \citet{JL_Torres2025} according to Gragera-Más et al. (in prep). The catalog combines \textit{Gaia} DR3 astrometry with radial velocities from RAVE DR6 \citep{Steinmetz2020}, APOGEE DR17 \citep{uf2022_APOGEE17}, GALAH DR4 \citep{Buder2025_GALAH4}, LAMOST DR11 \citep{Deng2012_LAMOST,Liu2020_LAMOST}, and DESI DR1 \citep{Koposov2026}, incorporating the latest survey releases and zero-point corrections relative to the \textit{Gaia} frame.

The complete encounter list is presented in Table~\ref{bp_close_enc}. For each perturber, \citet{JL_Torres2025} reconstructed the relative orbit with respect to BP through Monte Carlo integrations. In this work, we adopt the representative centroid solution in the space defined by the time and relative galactocentric positions and velocities of the encounter $(t,X,Y,Z,U,V,W)$ and derive the corresponding encounter time, minimum separation, and relative velocity, together with the estimated stellar mass of the perturber. 

We selected the perturbers whose reconstructed closest approaches occur after the present epoch (Table~\ref{bp_close_enc}). For each star, we adopted its present-day position and velocity relative to BP as the initial Cartesian state and integrated the system forward in time through the predicted encounter. All perturbers were included from the start of the simulation and evolved self-consistently with BP, its planets, and the planetesimal disk, allowing the close encounters to arise naturally during the integration rather than being imposed as instantaneous perturbations. Each perturber was removed 1~Myr after its closest approach, by which time even the slowest stars had receded by several parsecs and their dynamical influence on the system had become negligible.

\begin{figure}[ht]
\centering
\includegraphics[width=1.0\columnwidth]{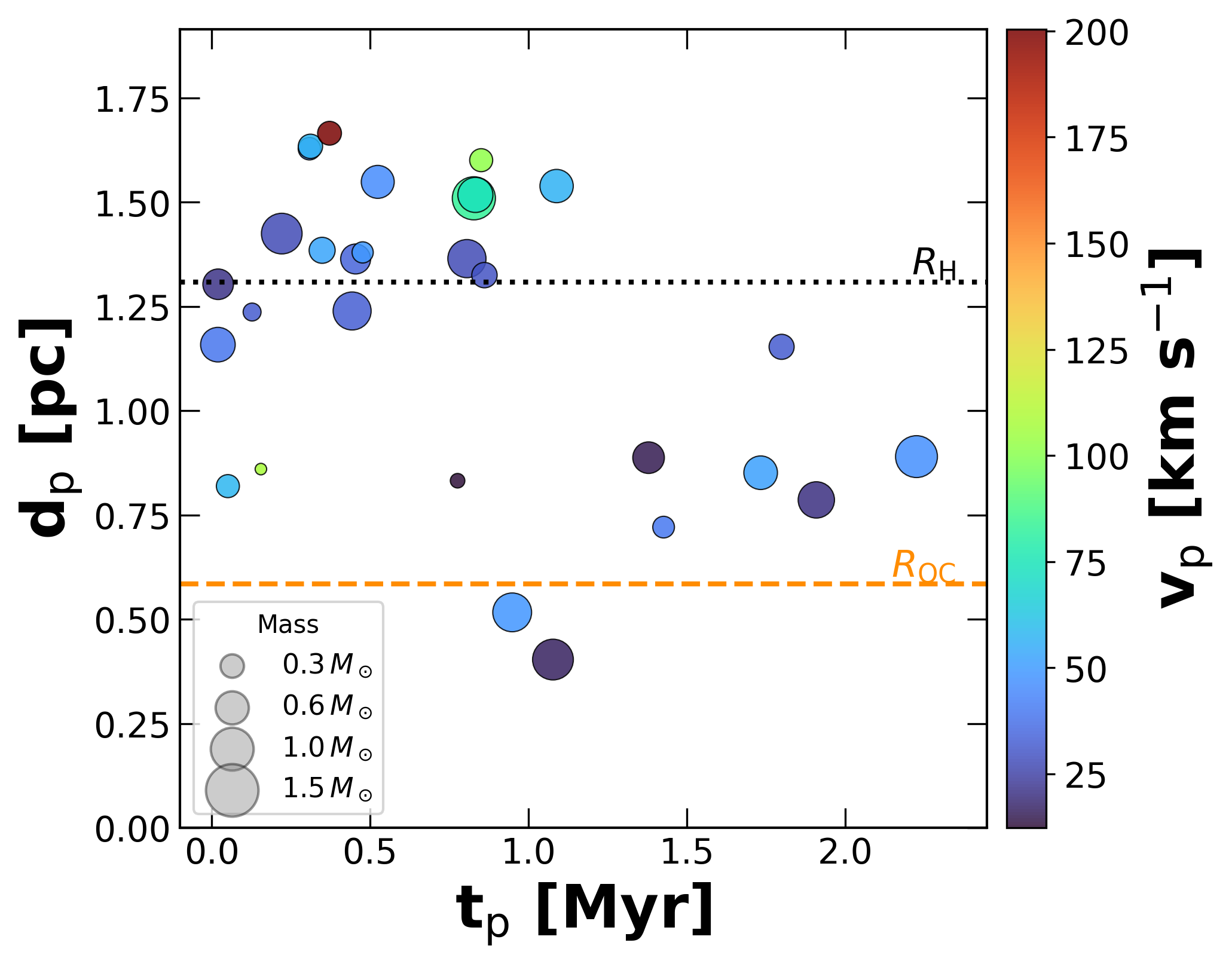}
\caption{Parameter space of the stellar encounters with BP considered in this work. The horizontal axis shows the time of closest approach relative to the present epoch, $t_{\rm p}$, and the vertical axis gives the corresponding minimum separation, $d_{\rm p}$. The color scale indicates the relative velocity at closest approach, $v_{\rm p}$, while the symbol size scales with the estimated mass of the perturber. The horizontal dashed and dotted lines indicate the fiducial Oort cloud radius (orange) and stellar Hill radius (black) adopted for the BP system, respectively.} 
\label{BP_encounters}
\end{figure}

Figure~\ref{BP_encounters} summarizes the parameter space occupied by the 29 encounters included in our model. Their minimum separations range from approximately $0.4$ to $1.75$ pc, while their relative velocities extend from a few tens to nearly $200~{\rm km\,s^{-1}}$. Most encounters occur within $2.5$ Myr of the present epoch. The majority are either relatively distant or fast, and only a small subset combines a low encounter velocity with a small minimum separation.

\begin{table*}[t]
\caption{Closest future stellar encounters with the Beta Pictoris system included in the simulations, ordered by closest-approach distance. The columns list the \textit{Gaia} DR3 source ID, stellar mass $M_\star$, time of closest approach relative to the present epoch $t_{\rm p}$, closest-approach distance $d_{\rm p}$, relative velocity at closest approach $v_{\rm p}$, and the estimated velocity impulse $\Delta v_\star$.}
\label{bp_close_enc}
\centering
\begin{tabular}{lccccc}
\hline\hline
\textit{Gaia} DR3 source ID &
$M_\star$ &
$t_{\rm p}$ &
$d_{\rm p}$ &
$v_{\rm p}$ &
$\Delta v_\star$ \\
&
($M_\odot$) &
(Myr) &
(pc) &
(km s$^{-1}$) &
(km s$^{-1}$) \\
\hline
5038817840251308288 & 0.90 & 1.078 & 0.403 & 16.27  & 1.183e-03 \\
6758141249403594112 & 0.82 & 0.949 & 0.516 & 48.63  & 2.800e-04 \\
4776148635544170752 & 0.26 & 1.427 & 0.721 & 39.54  & 7.703e-05 \\
6654230597344434560 & 0.72 & 1.909 & 0.786 & 19.92  & 3.937e-04 \\
4803556711646531840 & 0.29 & 0.051 & 0.819 & 58.23  & 5.200e-05 \\
4656782698309778432 & 0.11 & 0.776 & 0.832 & 12.33  & 9.584e-05 \\
3391664072252430976 & 0.61 & 1.733 & 0.851 & 51.55  & 1.202e-04 \\
5185493447310441728 & 0.07 & 0.156 & 0.860 & 107.63 & 6.697e-06 \\
6394330650108004992 & 0.54 & 1.380 & 0.887 & 14.69  & 3.574e-04 \\
3113120702787623808 & 0.95 & 2.225 & 0.890 & 46.69  & 1.972e-04 \\
1592423313280131200 & 0.34 & 1.800 & 1.153 & 31.73  & 8.029e-05 \\
4794632903476180096 & 0.65 & 0.020 & 1.158 & 39.02  & 1.238e-04 \\
4757687388639045504 & 0.17 & 0.128 & 1.237 & 31.81  & 3.814e-05 \\
5856411869205581568 & 0.78 & 0.444 & 1.239 & 32.19  & 1.692e-04 \\
5553110654636730496 & 0.51 & 0.021 & 1.303 & 20.90  & 1.601e-04 \\
3499296429434610304 & 0.35 & 0.861 & 1.325 & 29.63  & 7.662e-05 \\
4817064138977294592 & 0.49 & 0.455 & 1.364 & 34.08  & 9.018e-05 \\
2940796611884222208 & 0.79 & 0.806 & 1.365 & 27.62  & 1.806e-04 \\
2946531325238075776 & 0.24 & 0.477 & 1.380 & 49.50  & 3.072e-05 \\
2460983348274381696 & 0.37 & 0.349 & 1.385 & 53.47  & 4.275e-05 \\
5493588665684618752 & 0.90 & 0.221 & 1.425 & 27.39  & 1.980e-04 \\
4078432504018987904 & 1.01 & 0.828 & 1.509 & 82.92  & 6.947e-05 \\
3322241384816343040 & 0.67 & 0.833 & 1.518 & 73.46  & 5.158e-05 \\
6611488835158033920 & 0.60 & 1.089 & 1.539 & 56.82  & 5.934e-05 \\
6413811006857073536 & 0.59 & 0.524 & 1.549 & 45.77  & 7.160e-05 \\
3335744903730411520 & 0.29 & 0.851 & 1.601 & 101.44 & 1.524e-05 \\
3007559370624135424 & 0.27 & 0.309 & 1.628 & 44.53  & 3.252e-05 \\
2313022171603701888 & 0.32 & 0.312 & 1.634 & 57.84  & 2.923e-05 \\
4670295730560582784 & 0.31 & 0.372 & 1.666 & 200.45 & 7.927e-06 \\
\hline
\end{tabular}

\tablefoot{The estimated velocity change under the impulse approximation is
$\Delta v_\star \simeq 2GM_\star/(d_{\rm p}v_{\rm p})$, where $M_\star$
is the perturber mass, $d_{\rm p}$ is the closest-approach distance, and
$v_{\rm p}$ is the relative velocity at closest approach
\citep[e.g.,][]{Oort1950,Rickman1976,Dybczynski1994,Torres2019}.
This quantity provides an approximate measure of the absolute velocity
impulse associated with each flyby.}
\end{table*}


\section{Formation and Evolution of Minor-Body Populations}
\label{sec3}

To characterize the dynamical outcome of the simulations, we classify particles into five dynamical populations according to their orbital properties: scattered disk (SD), falling evaporating bodies (FEBs), Oort-cloud-like objects (OCOs), Transitional Interstellar Objects (TIOs), and interstellar objects (ISOs). Bound particles satisfy $a>0$ and $e<1$, whereas particles are classified as ISOs once they become gravitationally unbound with $e\geq1$. We define SD as particles with $a<1000$ au, $e>0.1$, and $q>3$ au, and FEBs as star-grazing objects with perihelion distances $q<q_{\rm FEB}$. We adopt $q_{\rm FEB}=0.4$ au, approximately corresponding to the calcium sublimation distance \citep[e.g.,][]{Beust1998,Beust2024,Jaworska2026}. 

To classify the distant bound populations, we introduce two characteristic spatial scales: a fiducial Oort-cloud radius, $R_{\rm OC}$, and the stellar Hill radius, $R_{\rm H}$. We estimate $R_{\rm OC}$ by scaling the characteristic outer radius of the Solar System Oort cloud with stellar mass, $R_{\rm OC}=R_{{\rm OC},\odot} (M_\star/M_\odot)^{1/3}$, where $R_{{\rm OC},\odot}=10^{5}\,\mathrm{AU}$ and $M_\star$ and $M_\odot$ are the masses of BP and the Sun, respectively \citep[e.g.,][]{Torres2020,OConnor2023b,Pham2024b}. This scaling is used as a fiducial reference rather than as a sharp physical boundary of an Oort cloud.

We define OCOs as bound particles with
$a_{\rm disk\,edge}<a<R_{\rm OC}$, where $a_{\rm disk\,edge}$ is the outer edge of the primordial disk. These objects represent the scattered population that begins to populate a proto-Oort cloud during the first 25~Myr of the system's evolution. The more weakly bound TIO population occupies the region
$R_{\rm OC}<a<R_{\rm H}$ \citep{Torres2019,Torres2020,Torres2025}. TIOs therefore occupy the weakly bound transition between the proto-Oort-cloud population and fully unbound ISOs. The stellar Hill radius sets the maximum spatial extent over which particles can remain gravitationally bound to BP. A bound object must satisfy $Q=a(1+e)\leq R_{\rm H}$, where $Q$ is the apocentre distance. For highly eccentric orbits with $e\approx1$, this corresponds approximately to $a_{\rm max}\simeq R_{\rm H}/2$. For BP, we adopt $R_{\rm OC}\approx1.2\times10^5~{\rm AU}$ and $R_{\rm H}\approx2.7\times10^5~{\rm AU}$. Table~\ref{table_fam} summarizes the dynamical populations adopted throughout this work.

\begin{table}
\centering
\caption{Definition of the dynamical populations adopted throughout this work: scattered disk (SD), falling evaporating bodies (FEBs), Oort-cloud-like objects (OCOs), Transitional Interstellar Objects (TIOs), and interstellar objects (ISOs). The columns list the population, semimajor axis ($a$), eccentricity ($e$), and perihelion distance ($q$), respectively. Here, $a_{\rm disk\,edge}$ denotes the outer edge of the primordial disk, $R_{\rm OC}$ the fiducial Oort-cloud radius, $R_{\rm H}$ the stellar Hill radius, and $q_{\rm FEB}=0.4$~au. A dash indicates that no additional constraint is imposed on that orbital element.}
\label{table_fam}
\begin{tabular}{lcccc}
\hline
Population & $a$\,[au] & $e$ & $q$\,[au] \\
\hline
FEBs      & $a>0$ & $e<1$ & $q<q_{\rm FEB}$ \\
SD        & $a<1000$    &   $0.1<e<1$ & $q>3$ \\
OCOs      & $a_{\rm disk\, edge}<a<R_{\rm OC}$ & $e<1$ & --- \\
TIOs      & $R_{\rm OC}<a<R_{\rm H}$ & $e<1$ & --- \\
ISOs      & --- & $e\geq1$ & --- \\
\hline
\end{tabular}
\end{table}

We now examine how the three-planet architecture reshapes the primordial disk and generates these populations over the first 25~Myr. We first characterize the planetary evolution and disk redistribution, then quantify the formation efficiencies and spatial structure of the resulting populations, and finally assess the influence of stellar encounters.

\subsection{Dynamical Evolution of Beta Pictoris' Planets}
\label{sec3.1}

\begin{figure}[ht]
\centering
\includegraphics[width=1.0\columnwidth]{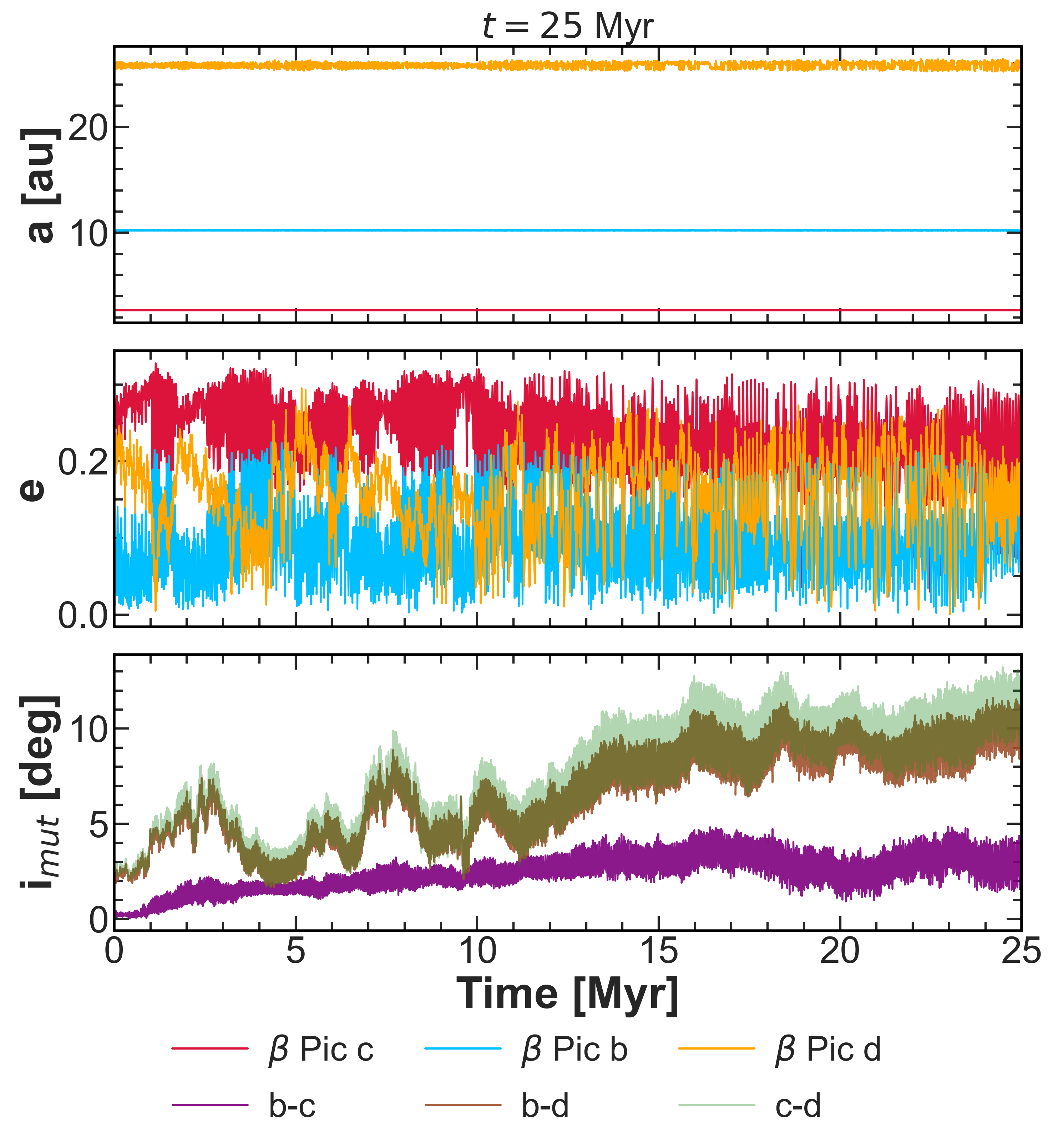}
\caption{Orbital evolution of Beta Pictoris planets over 25 Myr. Top: osculating semimajor axes ($a$) of $\beta$ Pic b, c, and d. Middle: corresponding eccentricities ($e$). Bottom: pairwise mutual inclinations ($i_{mut}$) calculated from the orbital angular-momentum vectors.}
\label{BP_planets}
\end{figure}

We first examine the dynamical evolution of the three known planets in the BP system, adopting the initial orbital configuration listed in Table~\ref{table_planets}. Figure~\ref{BP_planets} shows the evolution of their osculating Jacobian semimajor axes (top panel), eccentricities (middle panel), and pairwise mutual inclinations (bottom panel) throughout the 25~Myr integration.

All three planets remain gravitationally bound throughout the 25~Myr integration, and their hierarchical Jacobi semimajor axes undergo bounded oscillations rather than sustained migration. The semimajor axis of $\beta$ Pic c remains between $2.687$ and $2.691$~au, while that of $\beta$ Pic b spans $10.195$--$10.246$~au. The largest variation occurs for $\beta$ Pic d, which ranges from $25.182$ to $26.431$~au and ends the integration at $26.003$~au. The system therefore shows no evidence of large-scale orbital rearrangement over 25~Myr.

The eccentricities show stronger variability, indicating ongoing angular-momentum exchange between the planets. In particular, $\beta$ Pic b and d remain close to the 4:1 period commensurability: their median period ratio is $P_d/P_b=4.015$, and the pair remains within $2\%$ of the nominal commensurability for approximately $83\%$ of the integration. The mutual inclinations also vary systematically. The $\beta$ Pic b--c pair remains the most closely aligned, with a median mutual inclination of $2.40^{\circ}$ and a maximum of $4.86^{\circ}$. By contrast, the b-d and c-d pairs reach median mutual inclinations of $6.86^{\circ}$ and $7.47^{\circ}$, respectively, with maxima of $11.62^{\circ}$ and $13.22^{\circ}$. Their correlated evolution indicates that $\beta$ Pic b and c remain comparatively close to a common orbital plane, while most of the relative misalignment is associated with $\beta$ Pic d.

The possible dynamical role of an additional planet exterior to $\beta$ Pic b had already been explored before the discovery of $\beta$ Pic d. \citet{Lacquement2025} showed that the two-planet configuration could not reproduce the observed extent of the inner disk cavity, predicting an inner edge near $\sim28$~au rather than the observed $\sim50$~au. Their exploration of a hypothetical outer perturber favored dynamically viable solutions with eccentricities below $\sim0.4$. \citet{Gibbs2026} subsequently found that solutions with semimajor axes $\gtrsim30$~au are preferentially stable over 5~Myr, whereas the joint orbital analysis of \citet{Sutlieff2026} favors a semimajor axis near $26$~au and an orbital plane approximately aligned with those of $\beta$ Pic b and c. The newly discovered planet therefore occupies a region of parameter space already identified as dynamically relevant for shaping the inner edge of the debris disk, which we confirm in this study.

Our calculation is not a systematic stability survey of the allowed orbital parameter space. Instead, it demonstrates that the specific three-planet configuration adopted here remains bound and dynamically active over 25~Myr. The planetesimal disk therefore evolves under a planetary architecture that is stable on the age of the system while still providing the time-dependent perturbations that drive the disk evolution explored below.

\subsection{Planet--disk interaction and pathways to dynamical populations.}
\label{sec3.2}

\begin{figure}[ht]
\centering
\includegraphics[width=1.0\columnwidth]{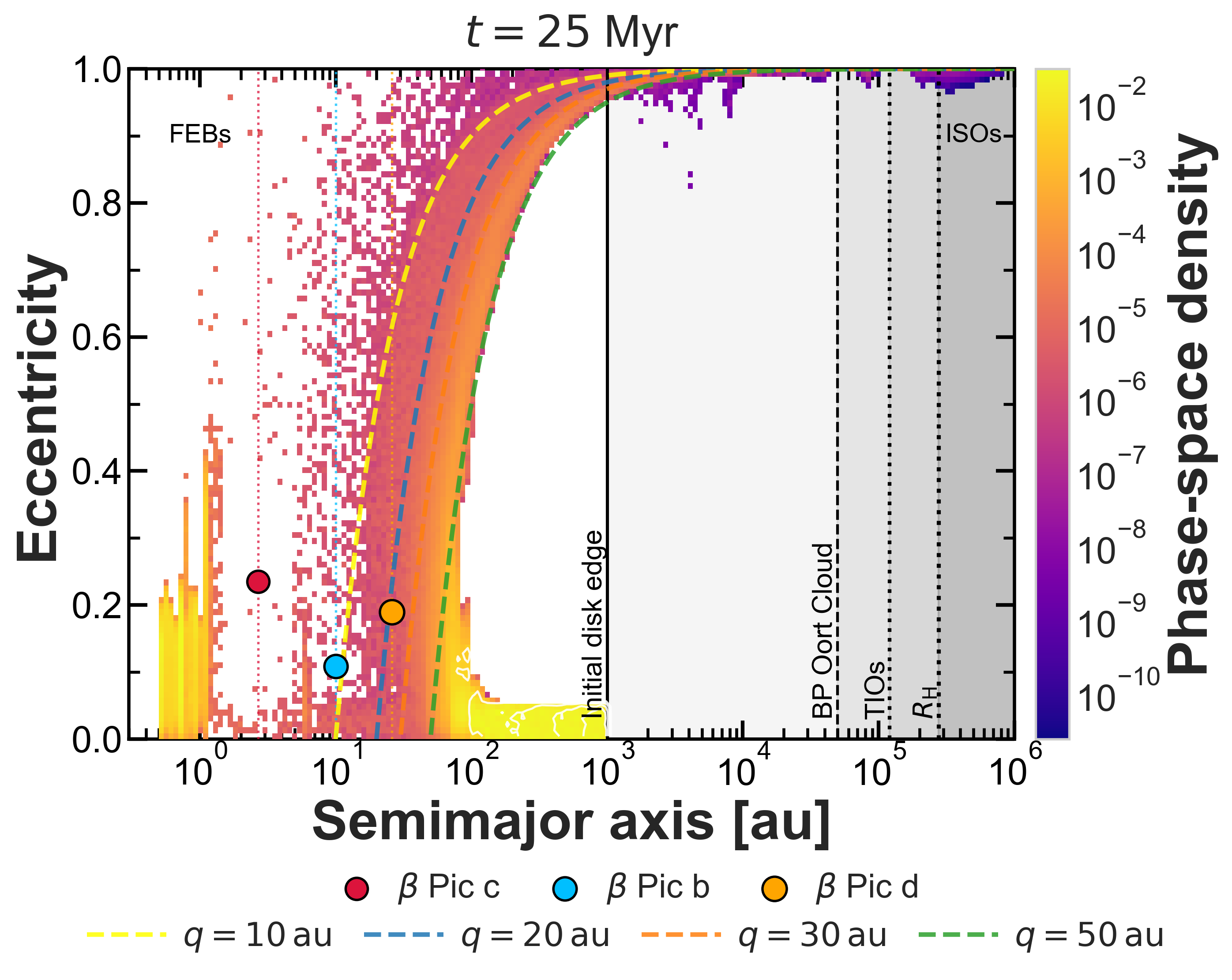}
\caption{Cumulative occupancy density of the planetesimal population in semimajor axis–eccentricity space. The stored-state counts are divided by the area of each (a,e) bin and normalized by the total number of states. Brighter regions therefore represent configurations occupied by more particles and/or for longer periods. Coloured circles and vertical dotted lines mark the planets, while the coloured dashed curves indicate constant perihelion distances of $q=10$, 20, 30, and 50 au. Black vertical lines mark the initial disk edge, the BP Oort cloud boundary, the transitional interstellar object (TIO) region, and the BP Hill radius. An animation of the planet--disk interaction can be found
\href{https://drive.google.com/drive/folders/1G7LQRl0L4j-SI8hv27NZTXLXZyKLEJ-W?usp=sharing}{online}.}
\label{BP_proto_OC}
\end{figure}

Having established that the three-planet architecture remains stable over 25~Myr, we now examine its interaction with the planetesimal disk. Figure~\ref{BP_proto_OC} shows the cumulative occupancy of the particles in semimajor axis ($a$)-eccentricity ($e$) space. Every stored orbital state contributes to the corresponding $(a,e)$ bin, so it reflects both the number of particles visiting a region and their residence time there. Curves of constant perihelion distance help distinguish planet-coupled scattering from dynamically detached evolution. Four principal structures emerge. 

The first is a compact inner population located interior to the orbit of $\beta$ Pic c. This feature is consistent with the inner ring identified by \citet{Beust2024} and \citet{Jaworska2026} as a potential source of FEBs. The FEB population in our simulations originates from the inner planetary region ($<20$~au), with $\beta$ Pic c acting as the main driver of this population. 

The second structure is an eccentricity tail extending outward from the planetary region, corresponding to the SD population and associated with the early formation of distant cometary populations \citep{PZ_ST_2021}. It is primarily populated by particles originating within approximately 70~au and follows trajectories with perihelia close to the planets. Along this branch, repeated planetary encounters change the orbital energy more efficiently than the perihelion distance, allowing particles to diffuse toward larger semimajor axes while remaining dynamically coupled to the planets. $\beta$ Pic d leaves a signature in this region, producing both eccentricity excitation and inclination growth. Although the locations of the 4:1--8:1 resonances coincide with substructure in the tail, the corresponding resonant angles show no persistent libration. Only isolated particles exhibit short episodes consistent with temporary 5:1 or 8:1 trapping (see Fig.~\ref{BP_proto_OC} and the
\href{https://drive.google.com/drive/folders/1G7LQRl0L4j-SI8hv27NZTXLXZyKLEJ-W?usp=sharing}{online animation}). 

The third structure is the sparse continuation of the scattering tail into the distant bound and unbound regions. Over the integration, $2.04\%$ of the initial particles enter the OCO region, while only $0.08\%$ reach the more distant TIO region. In comparison, $6.44\%$ of the particles become unbound and enter interstellar space. The fourth and final structure is the outer disk (150--1000~au), where particles do not undergo any classified transition and remain dynamically unevolved beyond approximately 325~au.

The resulting phase-space distribution traces a continuous scattering pathway from the inner disk to interstellar space. $\beta$ Pic d acts as the outermost planetary gateway, injecting particles into the high-eccentricity branch. Most of this material is eventually ejected or remains coupled to the planetary region, while a smaller fraction survives on distant bound orbits and contributes to the emerging Oort cloud.

\begin{figure*}[ht]
\centering
\includegraphics[width=1.0\textwidth]{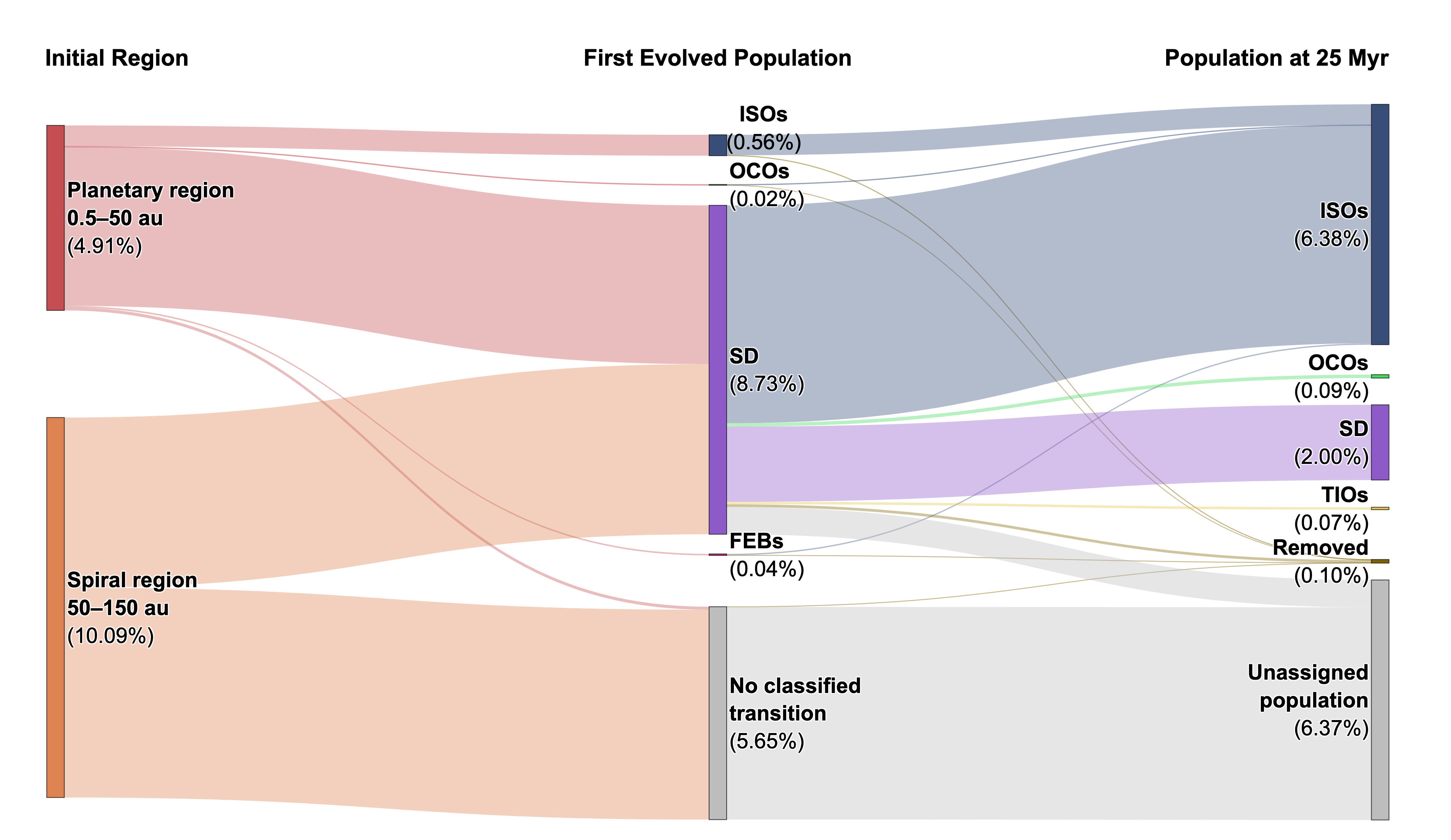}
\caption{Dynamical pathways of planetesimals originating within the dynamically active $0.5$--$150$~au disk. Flow widths trace particles from their initial radial region, through the first classified population entered, to the population they occupy at $25$~Myr. Percentages are given relative to the complete $20\,000$-particle sample. The brown node represents particles removed from the numerical integration after satisfying the adopted removal criteria. The outer disk ($150$--$1000$~au) is omitted from the diagram because no particle originating beyond $150$~au enters any of the adopted dynamical populations over 25~Myr. ISO, OCO, SD, TIO, and FEB denote interstellar objects, Oort-cloud-like objects, scattered-disk objects, transitional interstellar objects, and falling evaporating bodies, respectively. An interactive figure can be found
\href{https://drive.google.com/drive/folders/1G7LQRl0L4j-SI8hv27NZTXLXZyKLEJ-W?usp=sharing}{online}.}
\label{fig_sankey}
\end{figure*}

Figure~\ref{fig_sankey} connects the radial origin of the dynamically active particles with both the first dynamical population they enter and their population at 25~Myr. The scattered disk (SD) is the dominant first evolved population, encompassing $8.73\%$ of the complete disk. It is supplied by both the planetary region ($0.5$--$50$~au) and the spiral region ($50$--$150$~au), and subsequently acts as the main pathway toward more energetic outcomes. By contrast, only small fractions enter the ISO, OCO, and FEB populations directly as their first classified state.

The subsequent evolution of the scattered population is highly transient. Although $8.73\%$ of all particles enter the SD as their first evolved state, only $2.00\%$ remain in this class at 25~Myr. A large fraction of the material transported through the SD ultimately becomes unbound, with $6.38\%$ of the original disk occupying the ISO population at the final snapshot. The OCO and TIO populations contain only $0.09\%$ and $0.07\%$ of the original disk at 25~Myr, respectively. No particle satisfies the formal FEB criterion at the exact final snapshot, consistent with the highly transient nature of this population, while $0.10\%$ of the complete disk has been removed from the numerical integration due to our removal criteria.

At the first-classification stage, $5.65\%$ of the complete disk undergoes no classified transition. At 25~Myr, $6.37\%$ belongs to an unassigned population: these particles remain bound but do not satisfy any of the adopted population criteria at the final snapshot. This final category can include particles that experienced dynamical evolution earlier in the integration. The outer disk ($150$--$1000$~au) does not contribute to any of the classified dynamical populations over 25~Myr. Particles between $150$ and $325$~au may nevertheless undergo modest orbital evolution, whereas the disk beyond approximately $325$~au remains dynamically intact.

Overall, Fig.~\ref{fig_sankey} shows that the populations are connected stages of a common evolutionary sequence rather than independent outcomes. Planetary perturbations primarily transfer material from within $150$~au into the scattered disk, from which particles may remain on eccentric bound orbits, reach the distant OCO and TIO regions, or ultimately escape into interstellar space.

\subsection{Formation efficiency of the dynamical populations}
\label{sec3.3}
%


\begin{figure}[ht]
\centering
\includegraphics[width=1.0\columnwidth]{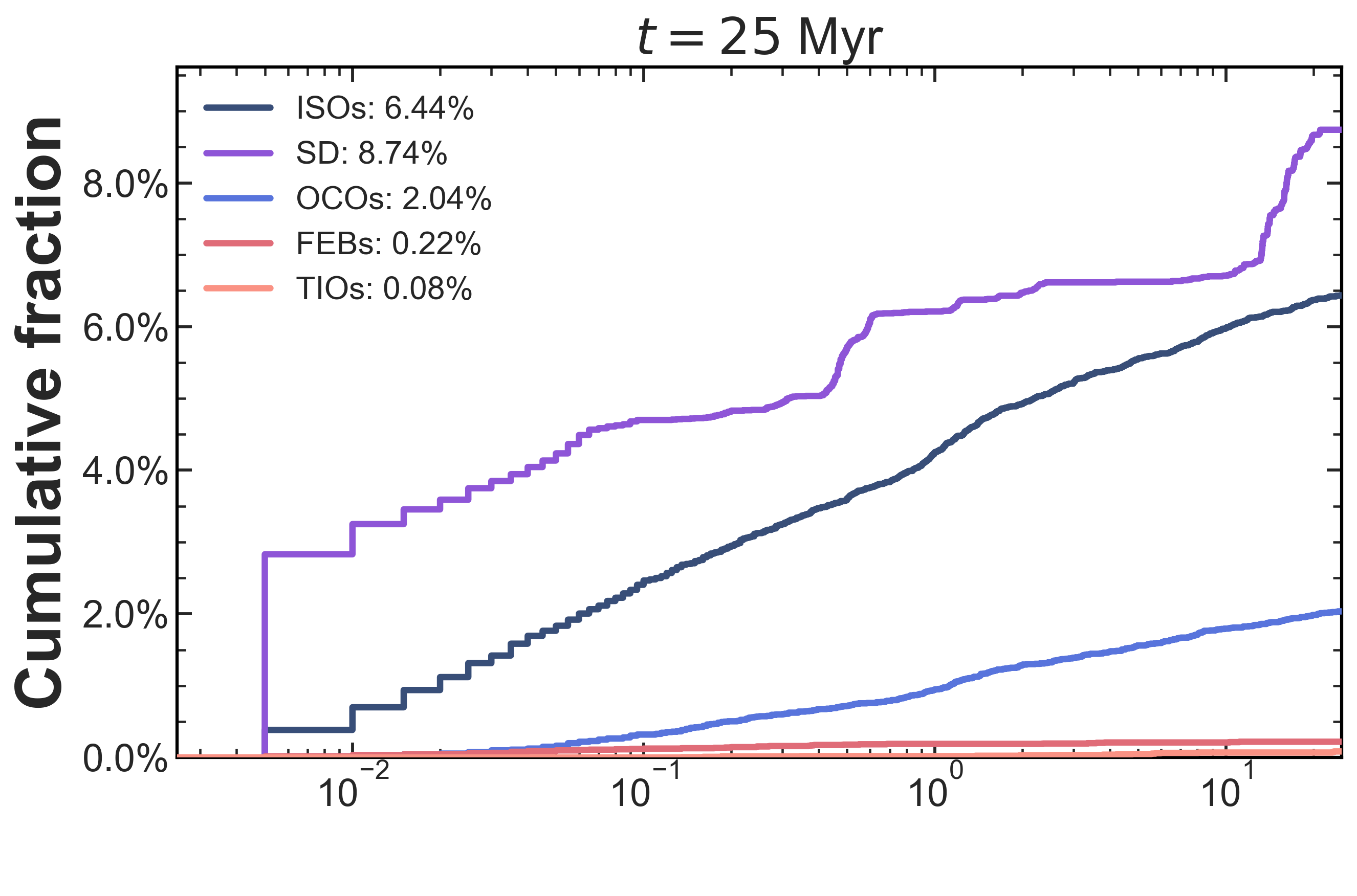}\\
\includegraphics[width=1.0\columnwidth]{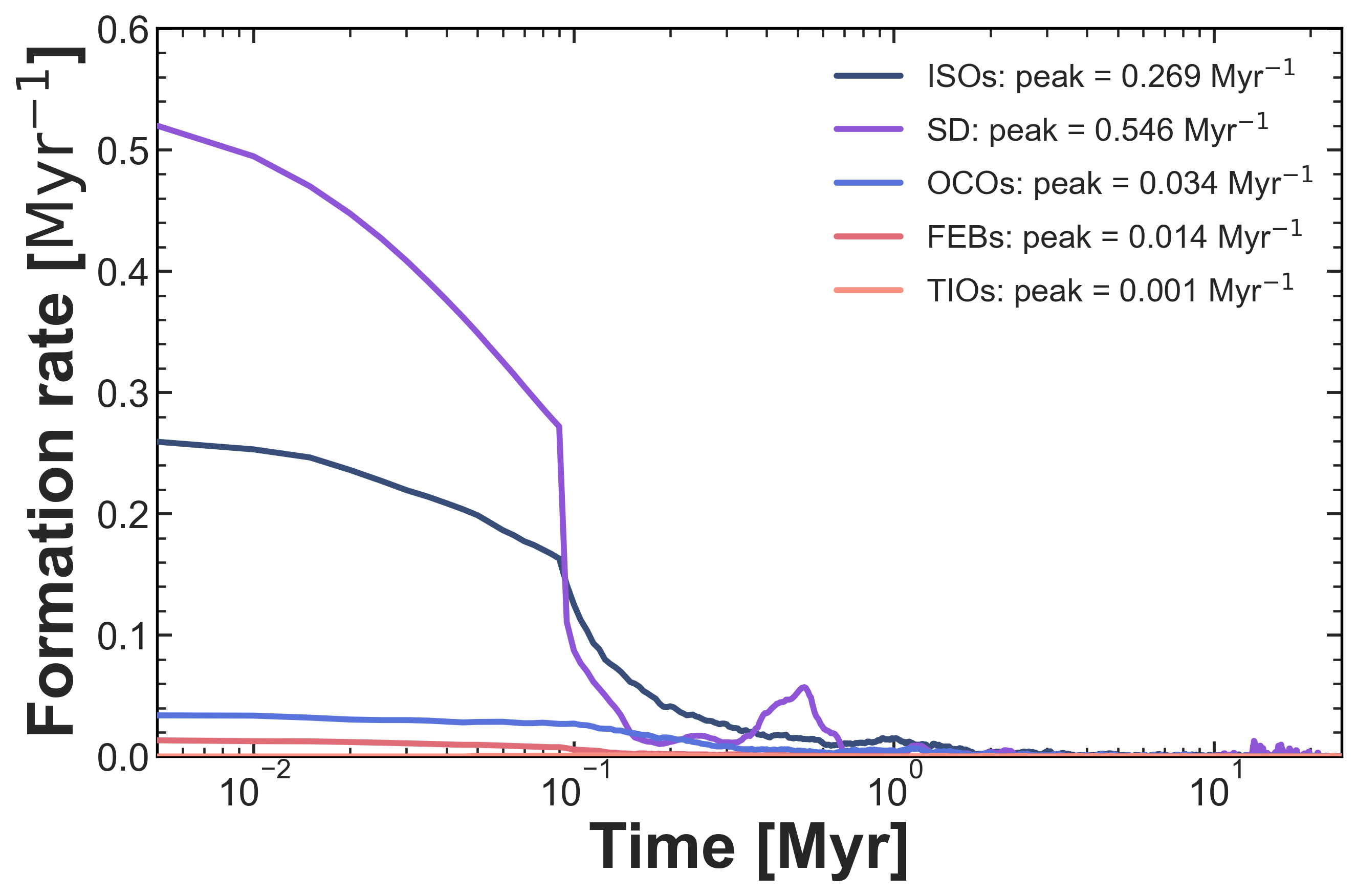}
\caption{Temporal evolution of the dynamical populations. The top panel shows the cumulative fraction of the initial particle population that enters each class at least once, while the bottom panel shows the corresponding first-entry rates. Colors denote interstellar objects (ISOs; dark blue), scattered disk (SD, purple), Oort-cloud-like objects (OCOs; light blue), falling evaporating bodies (FEBs; red), and transitional interstellar objects (TIOs; salmon).}
\label{fig_cdf}
\end{figure}

The phase-space analysis in Sec.~\ref{sec3.2} identifies the pathways through which planetesimals are redistributed. We now quantify their efficiencies and characteristic timescales. Figure~\ref{fig_cdf} shows the cumulative fraction of particles that have entered each population and the corresponding first-entry rate. Once a particle enters a population, it remains included in its cumulative count even if it subsequently leaves that state.

Figure~\ref{fig_cdf} shows that the scattered disk (SD) is the most frequently populated dynamical state, with $8.74\%$ of the initial disk entering this population during the integration. Its fractional first-entry rate initially reaches $0.546~{\rm Myr}^{-1}$, reflecting the rapid excitation of planetesimals in the dynamically active inner disk. However, only $2.00\%$ of the initial population remains in the SD at 25~Myr (Fig.~\ref{fig_sankey}), demonstrating that it acts primarily as an intermediate state through which particles evolve toward more distant bound or unbound trajectories.

The ISO population is the dominant escaping outcome. Its fractional first-entry rate initially reaches $0.269~{\rm Myr}^{-1}$ and declines sharply during the first $\sim0.1$~Myr before entering a slower, sustained phase. Particles ejected before and after this transition have significantly different initial semimajor-axis distributions, with later ejecta originating preferentially farther from the star. We therefore interpret the break as the rapid depletion of the most readily scattered inner population, followed by slower delivery from larger initial radii. Within the first Myr, $4.25\%$ of the disk becomes unbound, accounting for approximately $66\%$ of all ISOs produced over 25~Myr. The cumulative ISO fraction ultimately reaches $6.44\%$, while $6.38\%$ of the initial disk remains in the ISO population at the final snapshot.

Before becoming unbound, most confirmed ISOs pass through the planetary system. Of these objects, $88.56\%$, $42.38\%$, and $14.84\%$ reach perihelia within the orbital scales of $\beta$ Pic d, b, and c, respectively, while approximately $2\%$ enter the FEB region. For the present-day luminosity of BP, the nominal water-ice activity line lies near $8$--$10$~au. Approximately $40\%$ of the confirmed ISOs enter this region before ejection and may therefore experience water-ice sublimation and partial volatile depletion. The extent of this processing will depend on their sizes, thermal properties, number of perihelion passages, and residence times. Similar thermal processing has been proposed for Solar System Oort-cloud comets \citep[e.g.,][]{Gkotsinas2024}.

The ejected population has a median hyperbolic-excess velocity of $2.52~{\rm km\,s^{-1}}$. By 25~Myr, its logarithmic distance distribution peaks near $75$ pc and has a median distance of $60.4$ pc from BP. The typical ISO can therefore travel tens to hundreds of parsecs during the integration, after which its trajectory will be increasingly influenced by the Galactic potential and encounters with other stars \citep[e.g.,][]{Do2018,PZ_Torres2018,Hopkins2025ATLAS,XabiTorres2026}.

Transport into the BP Oort-cloud region is less efficient than scattering into the SD or interstellar space. Its fractional first-entry rate peaks at $0.034~{\rm Myr}^{-1}$, and its cumulative fraction reaches $2.04\%$ by 25~Myr. Only $0.09\%$ of the initial disk occupies the OCO region at the final epoch, indicating that this region also acts mainly as a transient intermediate state. The TIO population is rarer still: its first-entry rate peaks near $10^{-3}~{\rm Myr}^{-1}$, $0.08\%$ of the initial disk enters this weakly bound region during the simulation, and only $0.07\%$ remains there at 25~Myr.

FEB production is similarly rare and is strongly concentrated toward early times. Using the FEB criterion, $q<0.4$~au for bound orbits, $0.22\%$ of the initial population enters the FEB state during the integration, with a peak fractional first-entry rate of $0.014~{\rm Myr}^{-1}$. The activity declines rapidly after the first few Myr but does not cease completely: sporadic FEB configurations persist to late times, with the latest resolved episode occurring at $23.07$~Myr. No particle satisfies the FEB criterion at the exact 25~Myr snapshot, highlighting the strongly transient nature of this population.

The population histories therefore reveal two characteristic regimes. Rapid early scattering excites the SD and produces most of the ISOs and FEBs, whereas transport into the OCO and TIO regions continues over longer timescales. The SD and OCO populations are not independent terminal outcomes but intermediate stages in a connected scattering pathway. The three-planet architecture is consequently much more efficient at exciting and ejecting planetesimals than at retaining them on extremely distant bound orbits. Longer integrations including the Galactic tide and future stellar encounters will be required to determine the eventual fate of the weakly bound survivors.

\subsection{Structure of the evolved planetesimal disk}
\label{sec3.4}

\begin{figure*}
\centering
\includegraphics[width=\textwidth]{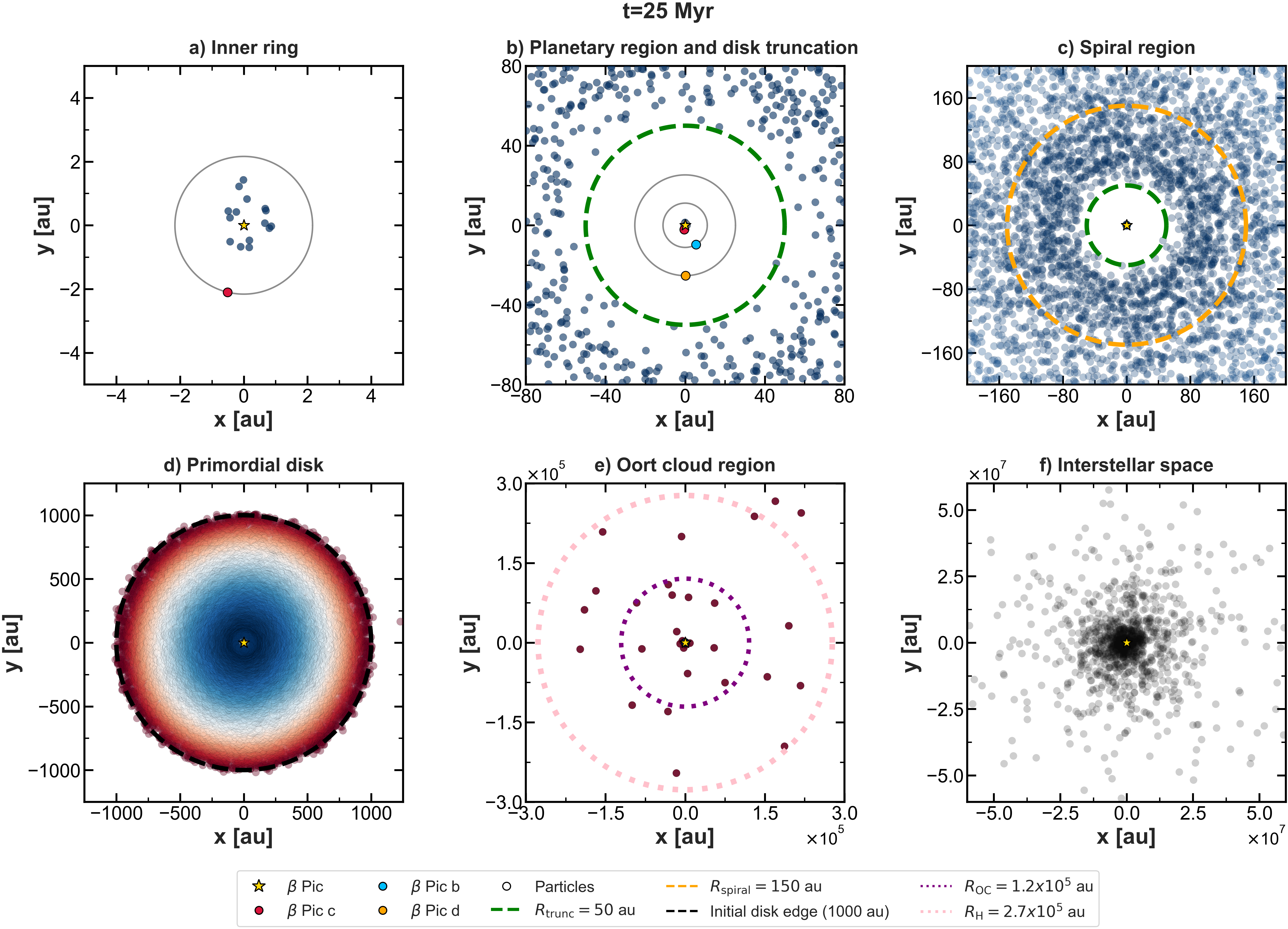}
\caption{Spatial structure of the $\beta$~Pictoris planetesimal disk after 25~Myr. The panels show progressively larger spatial scales, from the inner planetary region (a--b), the intermediate disk where a spiral-like non-axisymmetric structure develops at $\sim50$--$150$ au (c), the primordial disk  (d), the distant bound population (e), and the escaping population (f). Particles are colored according to their projected KDE density. The reference dashed circles indicate the main dynamical scales of the system: the disk truncation (green), the spiral region (orange), the initial disk edge (black), the BP Oort-cloud radius $R_{\rm OC}$ (purple), and the Hill radius $R_{\rm H}$ (pink). An animation of panel c can be found
\href{https://drive.google.com/drive/folders/1G7LQRl0L4j-SI8hv27NZTXLXZyKLEJ-W?usp=sharing}{online}.}
\label{fig_xy}
\end{figure*}

The planet-disk interactions described in the previous sections leave clear spatial signatures in the evolved planetesimal population. Figure~\ref{fig_xy} shows the projected $(x-y)$ distribution after 25 Myr, spanning scales from the inner planetary region to the escaping population. The final morphology shows that the three-planet architecture produces a strongly radius-dependent response: it clears and excites the inner disk, generates transient non-axisymmetric structure, and scatters a small fraction of planetesimals onto distant bound and unbound trajectories.

The innermost region (Fig.~\ref{fig_xy}a) retains a compact population interior to the orbit of $\beta$ Pic c. These particles correspond to the dynamically excited inner population capable of supplying star-grazing planetesimals, consistent with the mechanisms proposed by \citet{Beust2024} and \citet{Jaworska2026} and discussed in Sec.~\ref{sec3.2}. Moving outward, the planetary region (Fig.~\ref{fig_xy}b) shows strong depletion around the three giant planets. The surviving planetesimal disk develops a sharp inner edge near $50$ au, comparable to the observed and proposed inner boundary of the BP debris disk \citep{Augereau2001,Apai2015,Ballering2016,Han2026}. The same truncation is visible in the radial surface-density profile. The three-planet architecture therefore naturally clears the region interior to the outer disk, with $\beta$ Pic d defining the outermost planetary scattering boundary.

Beyond the cleared region, the surviving disk develops a transient, tightly wound one-armed spiral (Fig.~\ref{fig_xy}c and \href{https://drive.google.com/drive/folders/1G7LQRl0L4j-SI8hv27NZTXLXZyKLEJ-W?usp=sharing}{online animation}). We interpret this feature as an eccentricity spiral generated by planetary stirring. Secular perturbations cause planetesimals at different radii to precess at different rates, producing a radial gradient in the longitude of pericentre that winds an initially axisymmetric disk into a spiral \citep{Wyatt2005,Mustill2009,Kennedy2010}. This causes, after some delay, orbit-crossing in an initially dynamically-cold planetesimal disk, and can lead to high-velocity catastrophic collisions between planetesimals. A related mechanism was previously demonstrated for BP by \citet{Nesvold2015}, whose collisional model showed that $\beta$ Pic b can launch eccentricity and inclination waves and enhance collision velocities within a "stirring ring" at approximately $60$--$100$~au.

Our three-planet simulations reproduce this mechanism over a broader region, primarily among particles originating at $60\lesssim a_0\lesssim150$ au. During the spiral phase, neighboring radial intervals exhibit partial apsidal coherence, while their preferred longitudes of pericentre vary systematically with radius. Continued differential precession winds the pattern more tightly until phase mixing weakens its coherent spatial signature by 25 Myr, although the associated eccentricity and localized inclination excitation persist. We therefore interpret the feature as a transient planetary-stirring pattern rather than a long-lived density wave. 

Although our model does not include collisions, the resulting orbital excitation and orbit crossing could increase collision velocities and dust production in this region. The simulated spiral partially overlaps with the extended ``Cat's Tail'' observed in the BP disk \citep{Rebollido2024}, suggesting that planetary stirring may contribute to its formation. The BP disk also exhibits a strongly asymmetric CO distribution, including a prominent clump near $85$ au that may trace enhanced collisions among the bodies in the region \citep[e.g.,][]{Dent2014}. A direct connection cannot be established from the present simulations, however, because the observed structures involve dust and gas shaped by collisions, radiation pressure, and disk geometry, whereas our model follows massless planetesimals subject only to gravitational perturbations. Nevertheless, our results demonstrate that the current planetary architecture can generate transient, large-scale asymmetries at approximately $100$--$200$ au, highlighting the dynamical and potentially collisional importance of this region.

At the scale of the primordial disk (Fig.~\ref{fig_xy}d), the effect of the planetary system becomes progressively weaker with increasing distance. Beyond $a_0\simeq325$ au, particles remain bound, while the 90th-percentile orbital changes satisfy $|\Delta a|/a_0<0.02$, $|\Delta e|<0.01$, and $|\Delta i|<0.25^\circ$. The outer disk also preserves its initial vertical structure, with the median inclination changing only from $0.997^\circ$ to $0.999^\circ$. Thus, while the planets strongly restructure the inner few hundred au, the primordial population between approximately $325$ and $1000$ au remains essentially unchanged over 25 Myr. At larger scales, planetary scattering produces a sparse population of distant bound objects (Fig.~\ref{fig_xy}e). As quantified in Sec.~\ref{sec3.3}, some particles enter the region of the BP Oort cloud, $a_{\rm disk\,edge}<a<R_{\rm OC}$, while a still smaller fraction reaches the more weakly bound TIO region, $R_{\rm OC}<a<R_{\rm H}$. These populations represent the high-energy bound tail of the scattering distribution and should be regarded as the seed of an Oort cloud. Their long-term retention will depend on external perturbations capable of altering their perihelia and orbital energies.

Finally, the most energetic scattering events produce an escaping population that extends into interstellar space (Fig.~\ref{fig_xy}f). The spatial sequence shown in Fig.~\ref{fig_xy} therefore mirrors the dynamical pathways identified in Sec.~\ref{sec3.2}. The same planetary architecture that clears and excites the inner disk also scatters a small fraction of material toward distant bound orbits and ultimately into interstellar space. 

Overall, the response of the planetesimal disk is strongly dependent on radius. The planets dominate the inner few hundred au, where they produce clearing, dynamical excitation, and transient non-axisymmetric structure, while the primordial outer disk remains largely unaffected beyond $a_0\simeq325$ au. The population scattered to large distances provides the dynamical connection between the circumstellar disk and the wider stellar environment.

\subsection{The effect of stellar encounters}
\label{sec3.5}

The external stellar environment provides an additional source of perturbations that can modify the long-term evolution of planetary systems.  To quantify the recent external environment of the BP system, we include the Gaia-based stellar encounter catalog of \citet{JL_Torres2025}, updated according to Gragera-Más et al. (in prep). We select the 29 stellar passages with a probability greater than 90\% of passing within 2 pc of the system (Table~\ref{bp_close_enc}). The 29 selected perturbers are initialized from their present-day phase-space coordinates and integrated forward through their predicted encounters, as described in Sec.~\ref{sec2.2}.

The resulting evolution during the first 25 Myr remains dominated by the internal planet-disk interaction. The reconstructed flybys are generally characterized by relatively large impact parameters and high relative velocities, producing weak tidal perturbations compared with the repeated scattering generated by the giant planets. Even the strongest encounter in our sample (GDR3 5038817840251308288), with a closest approach of $d_{\rm p}=0.403$ pc and a relative velocity of $v_{\rm p}=16.27$ km s$^{-1}$, produces only a limited perturbation (Table~\ref{bp_close_enc}). Consequently, the cumulative fractions of all populations remain unchanged within the uncertainties introduced by the stellar encounters.

During the first 25 Myr, interactions with the giant planets dominate because most bodies remain comparatively close to the system, while the currently known flybys are neither sufficiently close nor slow to produce substantial perturbations. This conclusion nevertheless depends on the completeness of the encounter catalog: an as-yet unidentified close or slow encounter could significantly affect the disk, particularly its distant population. Stellar encounters are also expected to become increasingly important as planetary scattering populates weakly bound orbits, where even moderate perturbations can drive orbital diffusion, temporary capture, or re-injection into the inner system. Together with the Galactic tide, flybys may therefore regulate the later formation and long-term evolution of the BP Oort cloud \citep[e.g.,][]{Pfalzner2024,Torres2025}.

\section{Summary and Conclusions}
\label{conclusion}
In this work, we investigate the dynamical evolution of the Beta Pictoris system over 25~Myr, comparable to its estimated present age. Using the currently known three-planet architecture of $\beta$ Pic b, c, and d, we follow the evolution of a primordial planetesimal disk extending from 0.5 to 1000~au. The response of the disk is strongly radius-dependent: planetary perturbations efficiently restructure the inner $0.5$--$150$~au, while particles between approximately 150 and 325~au undergo only modest orbital evolution and do not enter any of the adopted dynamical populations. Beyond $a_0\simeq325$~au, the primordial population remains largely unchanged. Within the dynamically active inner region, the system develops distinct minor-body populations and disk structures whose main properties are summarized below.

\begin{itemize}

\item \textbf{Planetary architecture.}
The adopted three-planet configuration remains bound and dynamically active throughout the 25~Myr integration. The planetary semimajor axes remain confined near their initial values, while their eccentricities and mutual inclinations undergo coupled variations. In particular, $\beta$ Pic b and d remain close to the 4:1 period commensurability, with a median period ratio $P_d/P_b=4.015$.

\medskip

\item \textbf{Falling evaporating bodies (FEBs).} 
Planetary perturbations produce a small population of FEBs from the inner planetary system, primarily during the earliest stages of the evolution. The simulations retain a compact inner population interior to $\beta$ Pic c, consistent with the source region proposed in previous models. The inclusion of the newly discovered $\beta$ Pic d does not qualitatively alter this pathway: FEB production remains confined to the inner planetary system, with $\beta$ Pic c acting as the dominant driver of the star-grazing population. Although FEB activity is strongly concentrated at early times, sporadic star-grazing configurations persist to late epochs in the integration.

\medskip

\item \textbf{Disk truncation and planetary scattering.}
The three-planet architecture efficiently clears the planetary region and produces a surviving disk edge near $50$ au, consistent with the observed inner boundary of the outer debris disk. $\beta$ Pic d defines the outermost planetary scattering boundary and feeds a high-eccentricity branch extending outward from the planetary system. Substructure along this branch is associated with short episodes of temporary resonant trapping, although no persistent resonance is identified after 25 Myr.

\medskip

\item \textbf{Scattered disk.}
Planetary scattering produces a substantial scattered-disk population, with $8.74\%$ of the initial disk entering this state during the first 25 Myr. Its evolution is highly transient: only $2.00\%$ of the initial population remains in the scattered disk at the end of the integration. The scattered disk therefore acts primarily as an intermediate dynamical reservoir, linking the planet-coupled disk to more distant bound populations and ultimately to interstellar ejection.

\medskip

\item {\bf Transient spiral structure.}
Planetary stirring generates a tightly wound, one-armed spiral, primarily among particles originating at $60\lesssim a_0\lesssim150$ au. Differential apsidal precession progressively winds the pattern, while phase mixing weakens its coherent morphology by 25 Myr, although the underlying orbital excitation persists. The spiral overlaps the inner region of the observed Cat's Tail and encompasses the prominent CO clump near $\sim85$ au, suggesting that planetary stirring may contribute to the dynamical and collisional activity in this part of the disk. Establishing a direct connection, however, requires models that include collisions, radiation pressure, and the observed disk geometry.

\medskip

\item \textbf{BP Oort cloud.}
A smaller fraction of scattered planetesimals reaches distant bound orbits. Approximately $2.04\%$ of the adopted ensemble enters the fiducial BP Oort-cloud region during the integration, while only $0.08\%$ reaches the more weakly bound TIO region. Only a small fraction remains in either population at 25~Myr, indicating that these distant reservoirs are highly transient at the present stage of the system's evolution. They constitute the high-energy bound tail of the scattering distribution and should be interpreted as the seed of a future exo-Oort-cloud analog. Their subsequent retention, perihelion evolution, and orbital isotropization will depend on stellar encounters and the Galactic tidal field.

\medskip

\item \textbf{Interstellar objects (ISOs).}
Planetary scattering preferentially ejects material rather than retaining it on extremely distant bound orbits. By 25~Myr, $6.44\%$ of the adopted particle ensemble has become unbound, with approximately $66\%$ of these ISOs produced within the first Myr. Most experience substantial inward scattering before escape: $88.56\%$, $42.38\%$, and $14.84\%$ reach perihelia within the orbital scales of $\beta$ Pic d, b, and c, respectively, and about $40\%$ enter the nominal water-ice activity region. The escaping population may therefore be significantly thermally processed before entering interstellar space. The median hyperbolic-excess velocity is $2.52~{\rm km\,s^{-1}}$.

\medskip

\item \textbf{Stellar encounters.}
The reconstructed stellar flybys included in the present calculation produce only weak perturbations compared with the repeated scattering generated by the planets and do not measurably alter the formation efficiencies of the minor-body populations during the first 25~Myr. This result should, however, be interpreted in the context of the encounter catalog, which is only complete over the most recent $\sim2.5$~Myr. Additional past encounters that are not represented in the present reconstruction, as well as future stellar flybys, could modify the evolution of the more distant and weakly bound populations. Their importance is therefore expected to increase once planetary scattering has populated these regions, where even modest external perturbations can alter orbital energies and perihelia.

\end{itemize}

Taken together, these results suggest that Beta Pictoris is being observed during an active phase in the assembly of its minor-body populations. Planetary scattering continuously transfers material through the scattered disk toward distant bound orbits and interstellar space, producing the seeds of an exo-Oort cloud and an escaping ISO population. The present-day architecture therefore captures an important stage in this ongoing evolution, although earlier planetary growth, migration, or scattering may also have contributed to the current distribution. As material reaches increasingly weakly bound orbits, stellar encounters and the Galactic tide are expected to become progressively more important. Beta Pictoris thus provides a rare view of a young planetary system in which distinct minor-body populations are still being formed, transformed, and redistributed.

\begin{acknowledgements}
ST, EV and JLGM acknowledge support from the \textit{DISCOBOLO} project funded by the Spanish Ministerio de Ciencia, Innovaci\'on y Universidades under grant PID2021-127289NB-I00 and from the Agencia Estatal de Investigación (AEI) through the Severo Ochoa Centre of Excellence accreditation awarded to the Instituto de Astrofísica de Canarias, grant CEX2025-001609-S, funded by MICIU/AEI/10.13039/501100011033. AJM acknowledges support from the Swedish National Space Agency (Career Grant 2023-00146) and from the Swedish Research Council (Project Grant 2022-04043). ST acknowledges the contribution of the IAC High-Performance Computing support team and hardware facilities to this research.
\end{acknowledgements}

\bibliographystyle{aa} 
\bibliography{biblio_ST}



\end{document}